# New measurements of excitation functions of $^{186}$W(p,x) nuclear reactions up to 65 MeV.

## Production of $^{178}$W/$^{178m}$Ta generator

F. Tárkányi[a], F. Ditrói[a*], S. Takács [a] , A. Hermanne [b]

[a] Institute for Nuclear Research (ATOMKI), Debrecen, Hungary
[b] Cyclotron Department, Vrije Universiteit Brussel, (VUB), Brussels, Belgium

**Abstract**

New experimental excitation functions for proton induced reactions on $^{nat}$W are presented in the 32- 65 MeV energy range. The cross sections for $^{nat}$W(p,xn)$^{186,184,183,182m,182g,181}$Re, $^{na}$W(p,x)$^{178}$W, $^{nat}$W(p,x)$^{183,182,180m,177,176,175}$Ta, $^{175}$Hf, $^{177}$Lu were measured via an activation method by using a stacked-foil irradiation technique and high resolution gamma-ray spectrometry. The results were compared with predicted values obtained with the nuclear reaction model code TALYS (results taken from TENDL-2014 and TENDL-2015 on-line library). Production routes of the medically relevant radionuclides $^{186}$Re, $^{178}$W → $^{178}$Ta and $^{181}$W are discussed.

Keywords:

* Corresponding author: ditroi@atomki.hu

1. **Introduction**

Activation cross sections of radio products of light charged particle induced reactions on tungsten are important for different applications. The radionuclides [186]Re, [178]W/[178]Ta and [181]W are medically relevant. Tungsten is widely used as target material for spallation sources and accelerator-driven systems for transmutation of nuclear wastes and in nuclear fusion applications. In the frame of our systematic study of activation cross sections of light charged particle induced reactions for different applications and for comparison with theoretical predictions, we already investigated deuteron induced activation data on W up to 50 MeV [1, 2] and proton induced reactions up to 34 MeV [3]. These earlier studies and the related compilations were included in the IAEA organized recommended data libraries for production of therapeutic medical radioisotopes [4], and fusion evaluated data libraries [5]. The upgrade of therapeutic medical isotope data library is in progress. It will include a comprehensive comparison of production routes of the [178]W/[178]Ta generator. The compilation of proton induced routes shows that above 40 MeV only few experimental data, mostly from the Hannover group, exist at present.

The experimental data are also important for testing different theoretical nuclear reaction model codes. A special problem for the codes is the description of production yields of isomers. Due to the important applications (and having the opportunity of using the Louvain la Neuve cyclotron) we decided to extend the energy range of our measurements up to 65 MeV proton energy. For some applications and for optimal comparison with the cross sections predicted by codes, results on individual isotopes, and hence reaction cross sections, are required. Nevertheless, data on natural elements are relevant for many practical applications, and the quality of the theory can be evaluated either in limited energy regions where only one reaction contributes or from weighted and summed reaction cross section data over wider energy ranges. Lower costs, more easily available target material, better knowledge of isotopic composition and easier target preparation for commercially available metal foils are advantages of measuring so called "elemental cross sections".

2. **Earlier experimental works**

Numerous earlier experimental works (cross sections and thick target yields) were found in the literature up to 100 MeV published by [3, 6-28]. All these data were measured by activation followed by HPGe gamma-spectrometry. Significant systematic differences in the measured data are noted, but the information on the experimental conditions and the data evaluation processes reported in the original publications didn't allow to determine all reasons of disagreements.

### 3. Experiment and data evaluation

The experimental techniques and data analysis process were similar or identical to those described by us in recent publications. Here we present shortly the most important factors and some details specific for the present experiment.

One stack was irradiated at an external beam line of the Cyclone 90 cyclotron of the Université Catholique in Louvain la Neuve (LLN) with a proton beam of 65 MeV primary energy and 35 nA current for 1 h. The stack contained a 19 times repeated sequence of W(21.3 $\mu$m), Al(250 $\mu$m), GaNi alloy (17.7 or 15.9 $\mu$m) electrodeposited on 25 $\mu$m Au backing foils, Cu(12.5 $\mu$m) and Al(250 $\mu$m). The energy range covered by the 19 W targets was 32-65 MeV.

The activity produced in the targets and monitor foils was measured non-destructively (without chemical separation) using two high resolution HPGe gamma-ray spectrometers. Four series of measurements were performed at 25 and 5 cm source -detector distances  and started at 12.1-24.7 h, 119.5-166 h, 6577-6818 h and 6535-6838 h after the end of bombardment (EOB). The evaluation of the gamma-ray spectra were made by both a commercial [29] and an interactive peak fitting code [30].

The cross sections were calculated from the well-known activation formula from the measured activity, particle flux and number of target nuclei as input parameters. Some of the radionuclides formed are the result of cumulative processes (decay of metastable states or parent nuclides contribute to the formation process). Naturally occurring tungsten is composed of 5 stable isotopes ([180]W -0.13 %, [182]W -26.3 %, [183]W -14.3 %, [184]W -30.67 %, [186]W -28.6 %) therefore in most cases so called elemental cross sections were deduced, supposing the W to be monoisotopic with the total number of target atoms being the sum of all stable isotopes.

The decay data were taken from the online database NuDat2 [31] and the Q-values of the contributing reactions from the Q-value calculator [32], both are presented in Table 1.

*Effective beam energy* and the energy scale were determined primary by a stopping calculation [33] based on estimated incident energy and target thickness and finally corrected [34] on the basis of the excitation functions of the $^{24}$Al(p,x)$^{22,24}$Na monitor reactions [35] simultaneously re-measured over the whole energy range. For estimation of the uncertainty of the median energy of the incoming particles in the target samples and in the monitor foils the cumulative errors influencing the calculated energy (incident proton energy, thickness of the foils, beam straggling) have been taken into account. The uncertainty on the energy is in the ± 0 .5 - 1.5 MeV range, increasing towards the end of stack.

*The beam intensity* (the number of the incident particles per unit time) was obtained preliminary through measuring the charge collected in a short Faraday cup and corrected on the basis of the excitation functions of the monitor reactions compared to the latest version of IAEA-TECDOC-1211 recommended database [35].

It should be mentioned that in few cases we could not find independent gamma-lines to assess the produced activity of an investigated radioproduct. In these cases, the contributions of the overlapping gamma-lines from the decay of the other radioproducts were subtracted, or proper cooling time was applied or the through the activity of the short-lived daughter isotope.

Table 1.

Decay characteristics of the investigated activation products and Q-values of contributing reactions in $^{nat}$W(p,x) reactions

| Nuclide Decay path Level energy | Half-life | $E_\gamma$ (keV) | $I_\gamma$ (%) | Contributing reaction | GS-GS Q-value (keV) |
|---|---|---|---|---|---|
| **186Re**<br>β⁻: 92.53%<br>EC: 7.47% | 3.7183 d | 137.157 | 9.47 | 186W(p,n) | -1362.5 |
| **184mRe**<br>IT: 74.5%<br>EC: 25.5%<br>188.0463 keV | 169 d | 104.7395 | 13.6 | 184W(p,n)<br>186W(p,3n) | -2265.19<br>-15211.07 |
| **184Re**<br>EC: 100 % | 35.4 d | 111.2174<br>792.067<br>894.760<br>903.282 | 17.2<br>37.7<br>15.7<br>38.1 | 184W(p,n)<br>186W(p,3n) | -2265.19<br>-15211.07 |
| **183Re**<br>EC: 100 % | 70.0 d | 162.3266<br>291.7282 | 23.3<br>3.05 | 183W(p,n)<br>184W(p,2n)<br>186W(p,4n) | -1338.35<br>-8750.01<br>-21695.89 |
| **182mRe**<br>EC: 98.2 %<br>β⁺: 1.7 %<br>0+X (2+) | 14.14 h | 470.26<br>835.98<br>894.85<br>900.80 | 2.02<br>0.46<br>2.11<br>0.36 | 182W(p,n)<br>183W(p,2n)<br>184W(p,3n)<br>186W(p,5n) | -3582.0<br>-9773.0<br>-17185.0<br>-30131.0 |
| **182gRe**<br>EC: 99.99 %<br>β⁺: 0.01 %<br>(7+) | 64.0 h | 130.81<br>152.43<br>156.39<br>169.15<br>191.39<br>229.32<br>286.56<br>351.07<br>1076.2<br>1231.0 | 7.5<br>8.5<br>7.2<br>11.4<br>6.7<br>25.8<br>7.1<br>10.3<br>10.6<br>14.9 | 182W(p,n)<br>183W(p,2n)<br>184W(p,3n)<br>186W(p,5n) | -3582.0<br>-9773.0<br>-17185.0<br>-30131.0 |
| **181Re**<br>EC: 99.996 %<br>β+: 0.004 % | 19.9 h | 360.7<br>365.5 | 20<br>56 | 182W(p,2n)<br>183W(p,3n)<br>184W(p,4n)<br>186W(p,6n) | -10580.4<br>-16771.2<br>-24182.9<br>-37128.8 |
| **178W**<br>EC: 100 % | 21.6 d | no γ | | 180W(p,p2n)<br>182W(p,p4n)<br>183W(p,p5n)<br>184W(p,p6n)<br>186W(p,p8n)<br>178Re decay | -15371.8<br>-30123.6<br>-36314.4<br>-43726.1<br>-56672.0<br>-20910.1 |
| **183Ta**<br>β-: 100 % | 5.1 d | 353.9904 | 11.2 | 184W(p,2p)<br>186W(p,2p2n)<br>183Hf decay | -7700.42<br>-20646.29<br>-21873.9 |
| **182Ta**<br>β-: 100 % | 114.74 d | 1121.290<br>1189.040<br>1221.395<br>1231.004 | 35.24<br>16.49<br>27.23<br>11.62 | 183W(p,2p)<br>184W(p,2pn)<br>186W(p,2p3n)<br>182Hf decay | -7222.93<br>-14634.59<br>-27580.48<br>-14233.54 |
| **177Ta**<br>EC:100 % | 56.56 h | 112.9<br>208.4 | 7.2<br>0.94 | 180W(p,2p2n)<br>182W(p,2p4n)<br>183W(p,2p5n)<br>184W(p,2p6n)<br>186W(p,2p8n)<br>177W decay | -21353.12<br>-36104.91<br>-42295.73<br>-49707.4<br>-62653.29<br>-24150.8 |
| **176Ta**<br>EC: 99.11 %<br>β+: 0.89 % | 8.09 h | 88.35<br>201.84<br>710.50<br>1159.30 | 11.9<br>5.7<br>5.4<br>24.7 | 180W(p,2p3n)<br>182W(p,2p5n)<br>183W(p,2p6n)<br>184W(p,2p7n)<br>186W(p,2p9n)<br>176W decay | -29776.1<br>-44527.9<br>-50718.7<br>-58130.4<br>........<br>-31282.2 |
| **175Ta**<br>EC: 99.6 %<br>β+: 0.40 % | 10.5 h | 207.4<br>266.9<br>348.5 | 14.0<br>10.8<br>12.0 | 180W(p,2p4n)<br>182W(p,2p6n)<br>183W(p,2p7n)<br>184W(p,2p8n) | -36804.2<br>-51555.9<br>-57746.8<br>-65158.4 |

| | | | | $^{175}$W decay | -40362.4 |
|---|---|---|---|---|---|
| **$^{180m}$Hf**<br>β⁻: 0.31 %<br>IT: 99.69 %<br>1141.573 *keV* | 5.47 h | 93.325<br>215.426<br>332.275<br>443.163<br>500.697 | 17.1<br>81.3<br>94.1<br>81.9<br>14.3 | $^{182}$W(p,3p)<br>$^{183}$W(p,3pn)<br>$^{184}$W(p,3p2n)<br>$^{186}$W(p,3p4n) | -13043.86<br>-19234.68<br>-26646.34<br>-39592.22 |
| **$^{175}$Hf**<br>EC: 100 % | 70 d | 343.40 | 84 | $^{180}$W(p,3p3n)<br>$^{182}$W(p,3p5n)<br>$^{183}$W(p,3p6n)<br>$^{184}$W(p,3p7n)<br>$^{186}$W(p,3p9n)<br>$^{175}$Ta decay | -33946.62<br>-48698.41<br>-54889.23<br>-62300.89<br><br>-8508.5 |
| **$^{177}$Lu**<br>β⁻: 100 % | 6.647 d | 112.9498<br>208.3662 | 6.17<br>10.36 | $^{180}$W(p,4p)<br>$^{182}$W(p,4p2n)<br>$^{183}$W(p,4p3n)<br>$^{184}$W(p,4p4n)<br>$^{186}$W(p,4p6n) | -19119.67<br>-33871.46<br>-40062.28<br>-47473.94<br>-60419.83 |

Isotopic abundance: $^{180}$W -0.13 %, $^{182}$W -26.3 %, $^{183}$W -14.3 %, $^{184}$W -30.67 %, $^{186}$W -28.6 %. When complex particles are emitted instead of individual protons and neutrons the Q-values have to be decreased by the respective binding energies of the compound particles: np-d, +2.2 MeV; 2np-t, +8.48 MeV; 2p2n-α, +28.30 MeV.

Decrease Q-values for isomeric states with level energy of the isomer.

## 4. Excitation functions

We deduced cross sections for the $^{nat}$W(p,xn)$^{186,184,183,182m,182g,181}$Re, $^{na}$W(p,x)$^{178}$W, $^{nat}$W(p,x)$^{183,182,177,176,175}$Ta, $^{180m,175}$Hf, $^{177}$Lu reactions. The excitation functions are shown in Figures 1-16. The numerical data are collected in Tables 2 and 3. The contributing reactions for the investigated products are collected in Table 1. The cross section data are compared with prediction obtained from the latest version of the TENDL library [36]. It is based TALYS code family (version 1.8) [37]. The calculations are based on default parameters, but in case of protons it includes adjusted results.

### 4.1 Production of $^{186}$Re

The radioisotope $^{186}$Re (3.7183 d) has a potential application in internal radiotherapy. It is formed on $^{nat}$W only directly by the $^{186}$W(p,n) reaction. Reaction cross sections can hence be derived from our measurements. As different production routes will be discussed in more detail in the section "Production routes of medically relevant radioisotopes", we present in the figure all the available experimental and theoretical data. Our measurements form a good extension to higher energies of cross sections published by [19, 20, 26] up to 30 MeV and are in agreement with [14] between 50 and 60 MeV.

The experimental data are a factor of two higher at the maximum of the excitation function than values tabulated in TENDL libraries. The overall shape is well described.

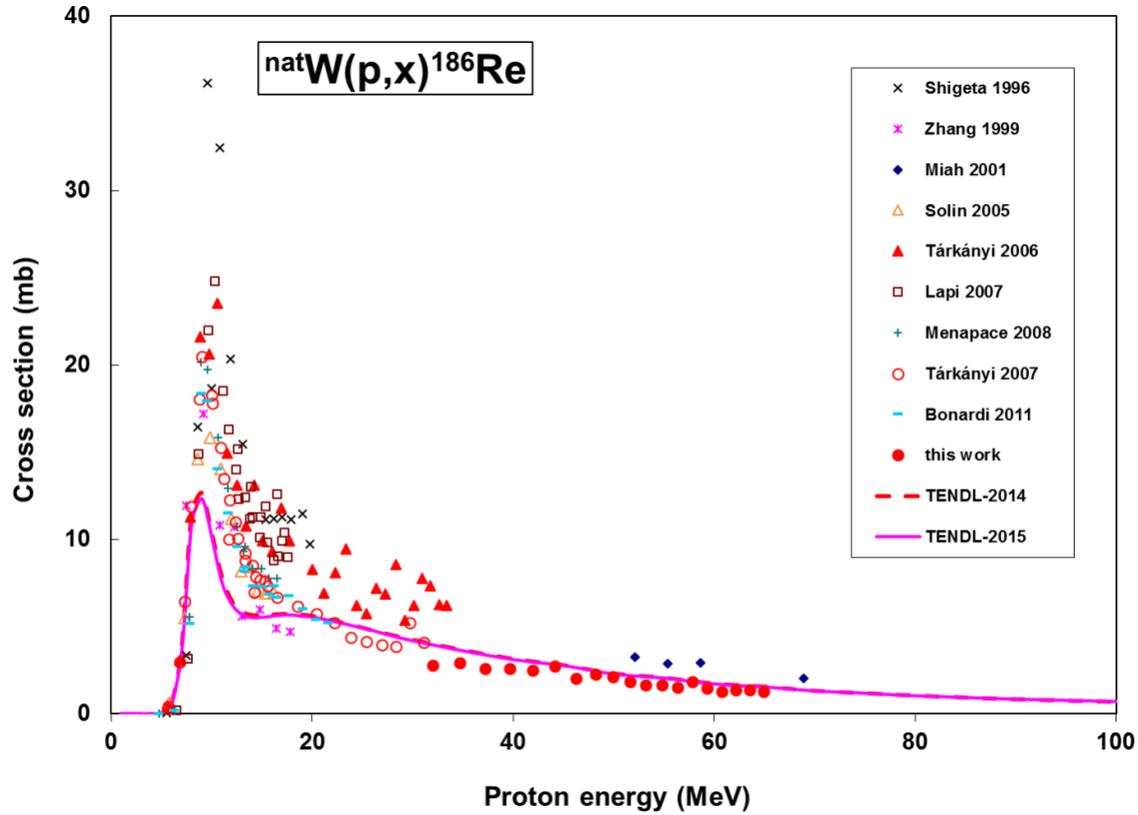

Fig. 1. Experimental excitation function for the $^{nat}W(p,x)^{186}Re$ reaction and comparison with literature values and theoretical code calculations

**4.2    Production of $^{184m}Re$ and $^{184g}Re$**

The radionuclide [184]Re has two observable isomeric states. A 169 d half-life isomeric state that decays for 74% by IT to the shorter- lived ground state (35.4 d). The ground state has practically no independent gamma-lines as its major gamma-lines are also occurring with low intensity in the decay of the metastable state. The cross section of the isomeric state was hence determined via its independent gamma-lines from spectra measured after complete decay of the ground state. The cross sections of the ground state are based on gamma spectra measured 5-10 days after EOB. Both contributions of decay of the isomeric state (common $\gamma$-lines and isomeric transition) was corrected for. Our measurements for the metastable state  are in agreement with the high energy experiments of [14, 16]. No experimental results in the expected region of the maximum cross section are available. TENDL overestimates the formation cross sections of the isomeric state (Fig. 2). For the ground state an acceptable agreement with [6, 14, 17] is seen in Fig.3 while here TENDL-2014 underestimated the maximum value.

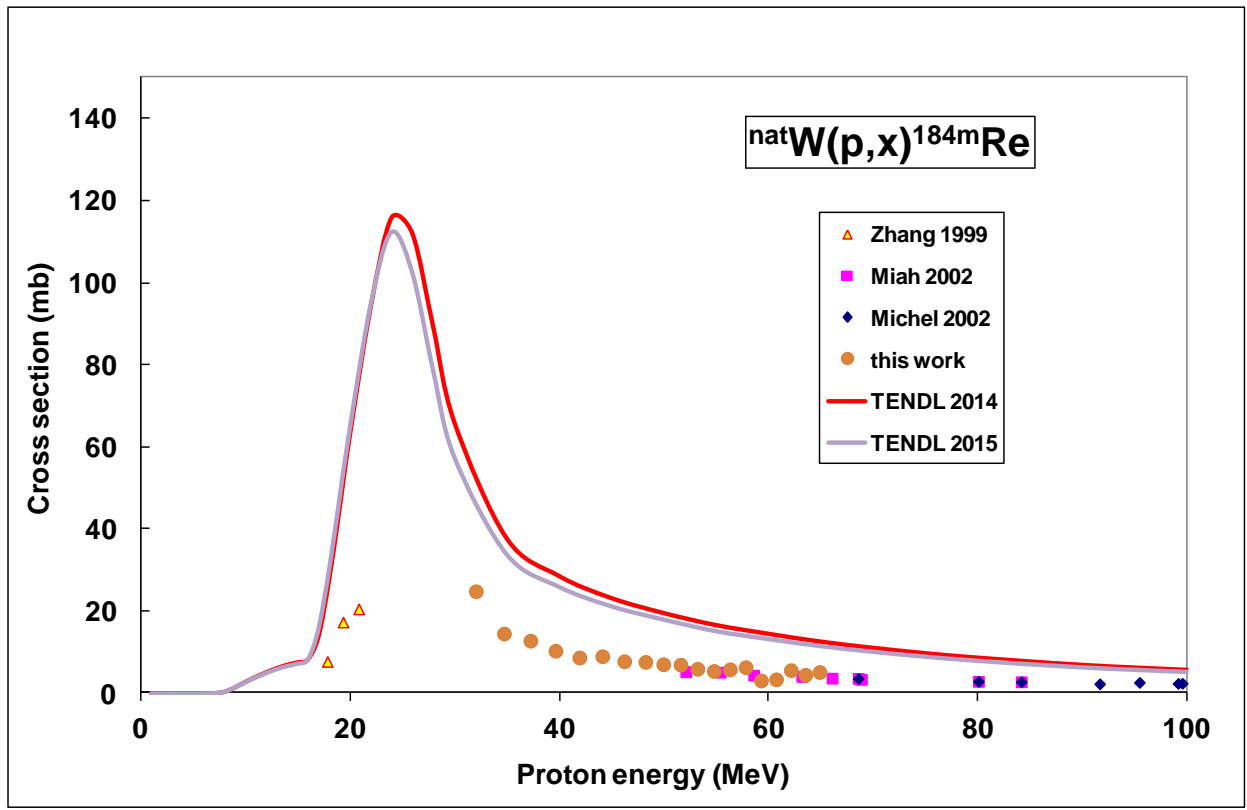

Fig. 2. Experimental excitation function for the natW(p,x)184mRe reaction and comparison with literature values and theoretical code calculations

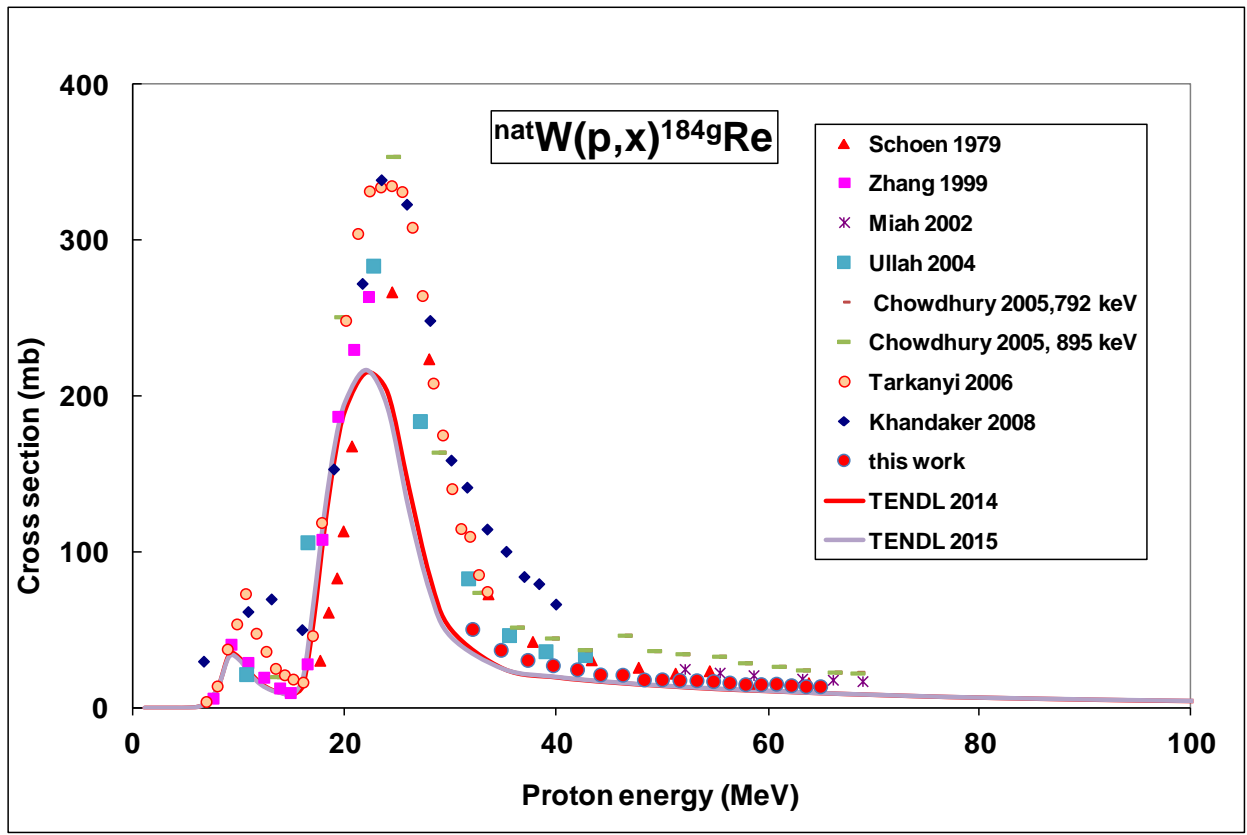

Fig. 3. Experimental excitation function for the $^{nat}$W(p,x)$^{184g}$Re reaction and comparison with literature values and theoretical code calculations

### *4.3    Production of $^{183}$Re*

According to Fig. 4 the main contributions to formation of $^{183}$Re (70.0 d) are the $^{184}$W(p,2n) and $^{186}$W(p,4n) reactions (see Table 1 for thresholds). The earlier experimental data of [6, 9, 21] are discrepant with the other studies. Our measurements form a good extension to higher energies of cross sections published by [3, 25] up to 40 MeV and are in agreement with [14, 16]. The agreement between our new data and the theoretical cross sections values is acceptable.

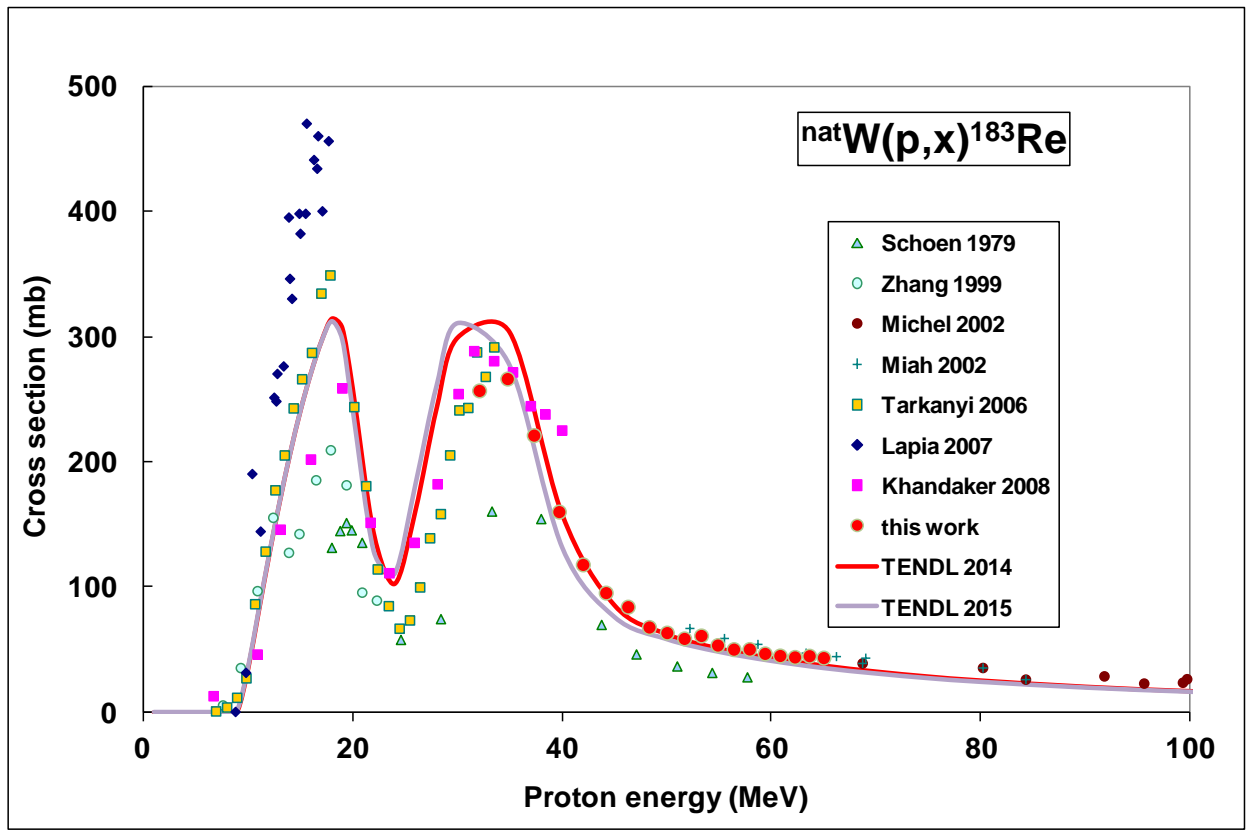

Fig. 4. Experimental excitation function for the $^{nat}W(p,x)^{183}Re$ reaction and comparison with literature values and theoretical code calculations

### 4.4    Production of $^{182m}$Re and $^{182g}$Re

The radionuclide $^{182}$Re has two observable isomeric states. The 64.0 h ground-state and the 12.7 h half-life higher laying isomeric state are decaying independently to stable $^{182}$W. Multiple reactions on the different stable W isotopes contribute to the formation. Our data for the metastable state $^{182m}$Re fit well to the earlier experimental data obtained at lower energy and with the high energy experiment of [38]  while [14] is discrepant (Fig. 5). The TENDL predictions are too low.

For the ground state a general good agreement is found with literature in the overlapping energy region and with high energy values of [14, 16]. The theoretical description is better for the ground state (Fig. 6), but the overestimation of the first maxima is significant.

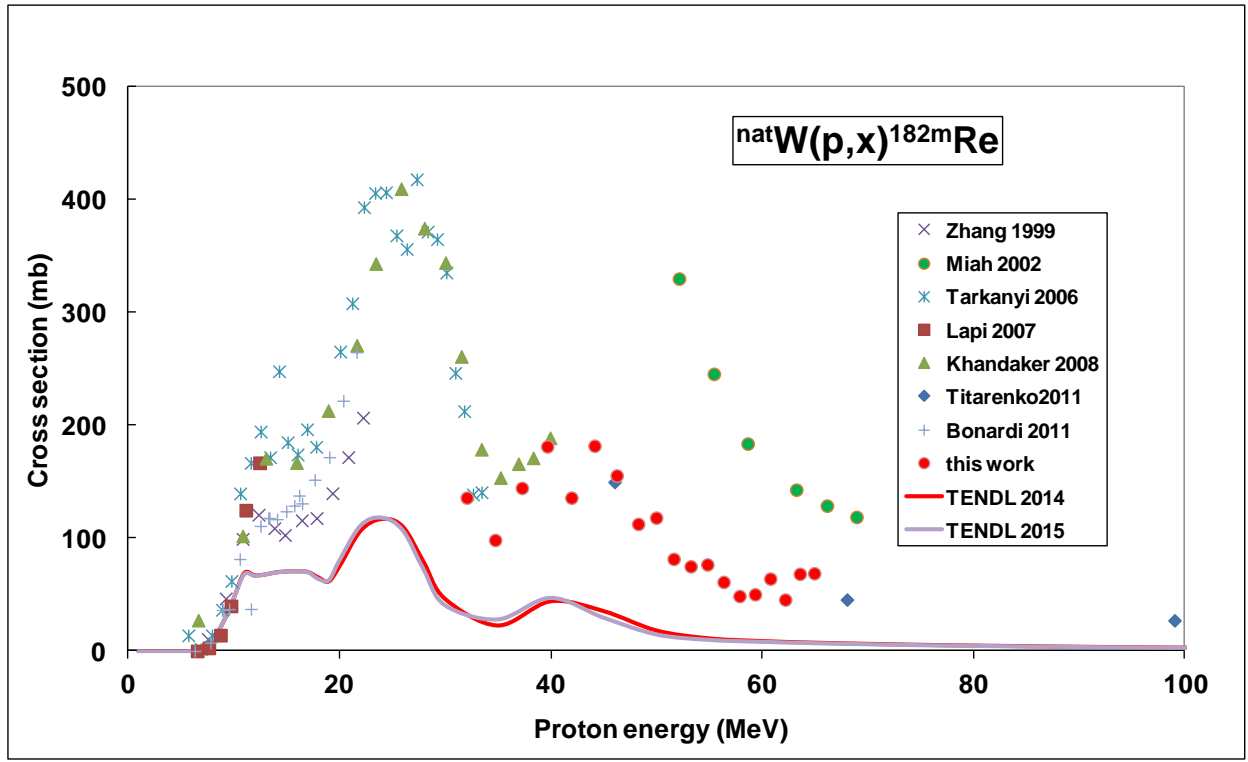

Fig. 5. Experimental excitation function for the $^{nat}$W(p,x)$^{182m}$Re reaction and comparison with literature values and theoretical code calculations

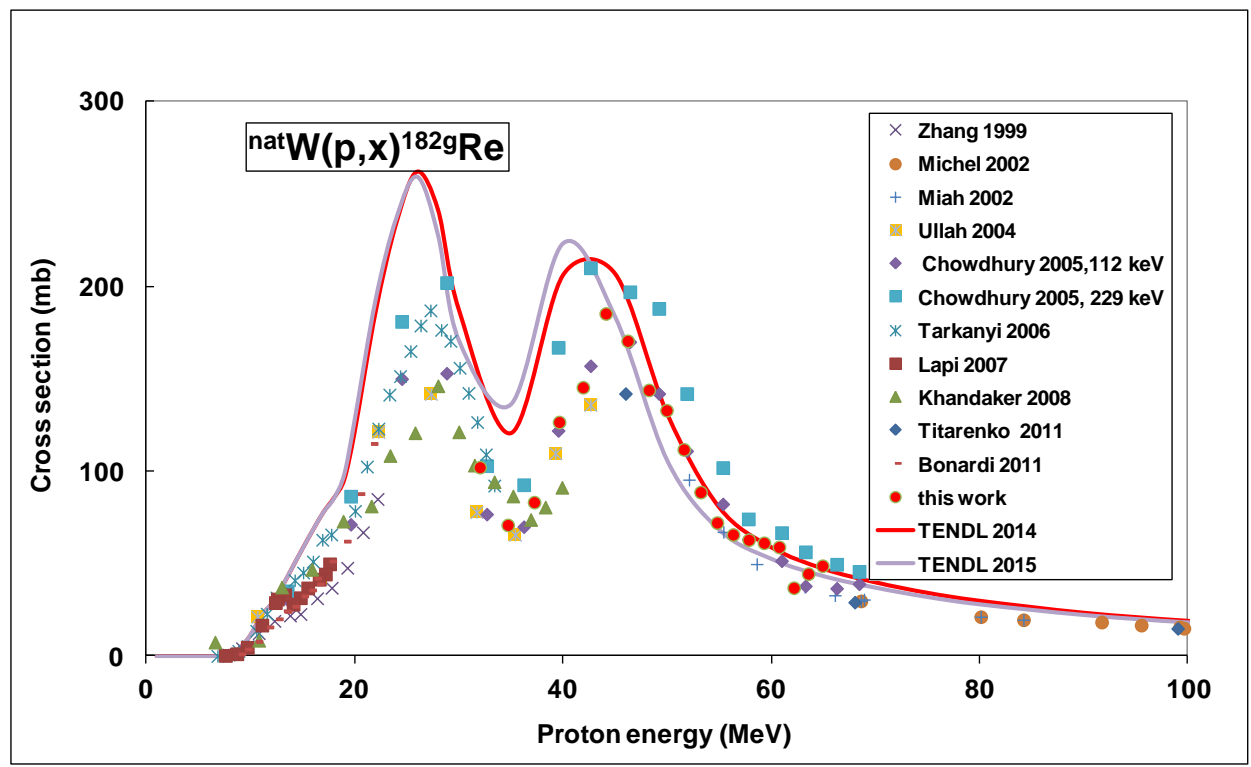

Fig. 6. Experimental excitation function for the $^{nat}W(p,x)^{182g}Re$ reaction and comparison with literature values and theoretical code calculations

### 4.5 Production of $^{181}Re$

The earlier experimental data for production of $^{181}Re$ (19.9 h) show clearly the multiple formation and our new values are in good agreement in the overlapping energy region (Fig. 7) (except for data in the MSc work of [18]. The TENDL predictions are acceptable, except above 50 MeV, where an underestimation by the theory can be observed.

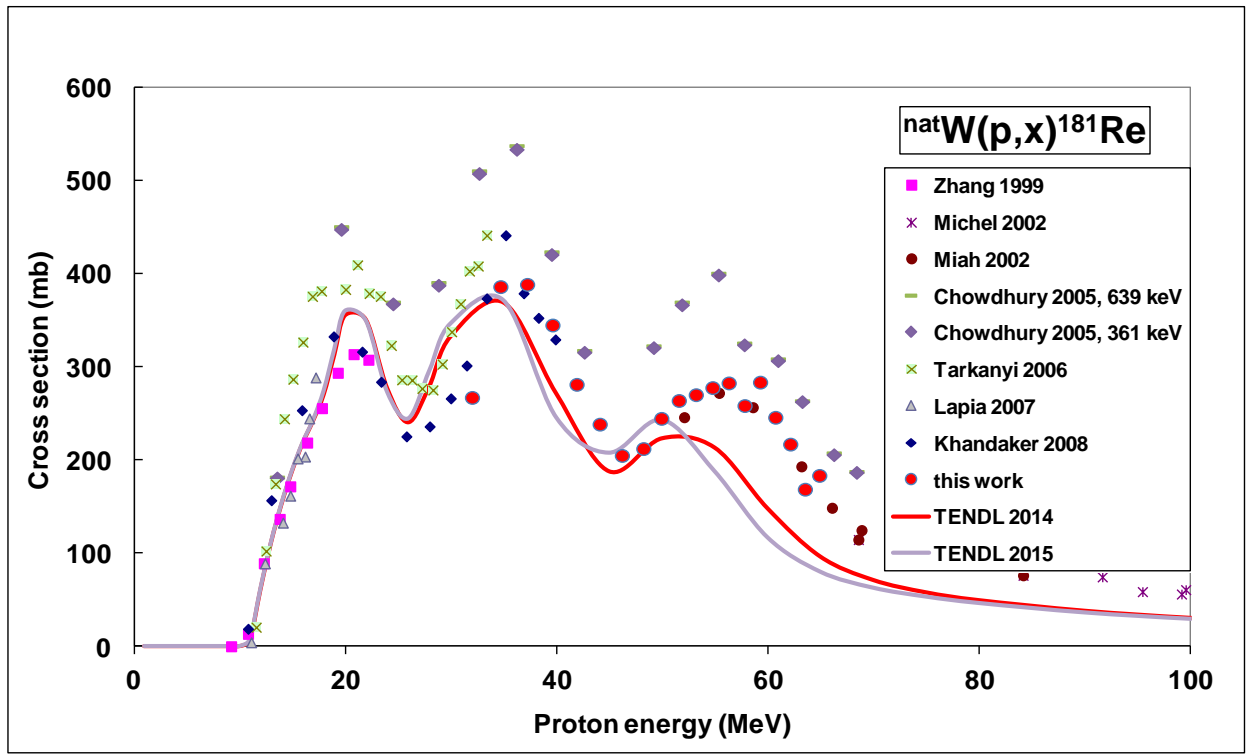

Fig. 7. Experimental excitation function for the $^{nat}$W(p,x)$^{181}$Re reaction and comparison with literature values and theoretical code calculations

## 4.6   Production of $^{178}$W

The radionuclide $^{178}$W (32.508 d) is produced directly via (p,pxn) reactions and from the decay of shorter-lived $^{178}$Re (13.2 min) parent. The experimental data shown in Fig. 8 are cumulative as they were obtained from spectra measured after the complete decay of $^{178}$Re. The agreement between the experimental and theoretical data is acceptable. The predicted curve for $^{180}$W(p,p3n) is also presented for discussion later, although, because of the very low abundance (< 0.2%) it plays no observable role in the excitation function on $^{nat}$W.

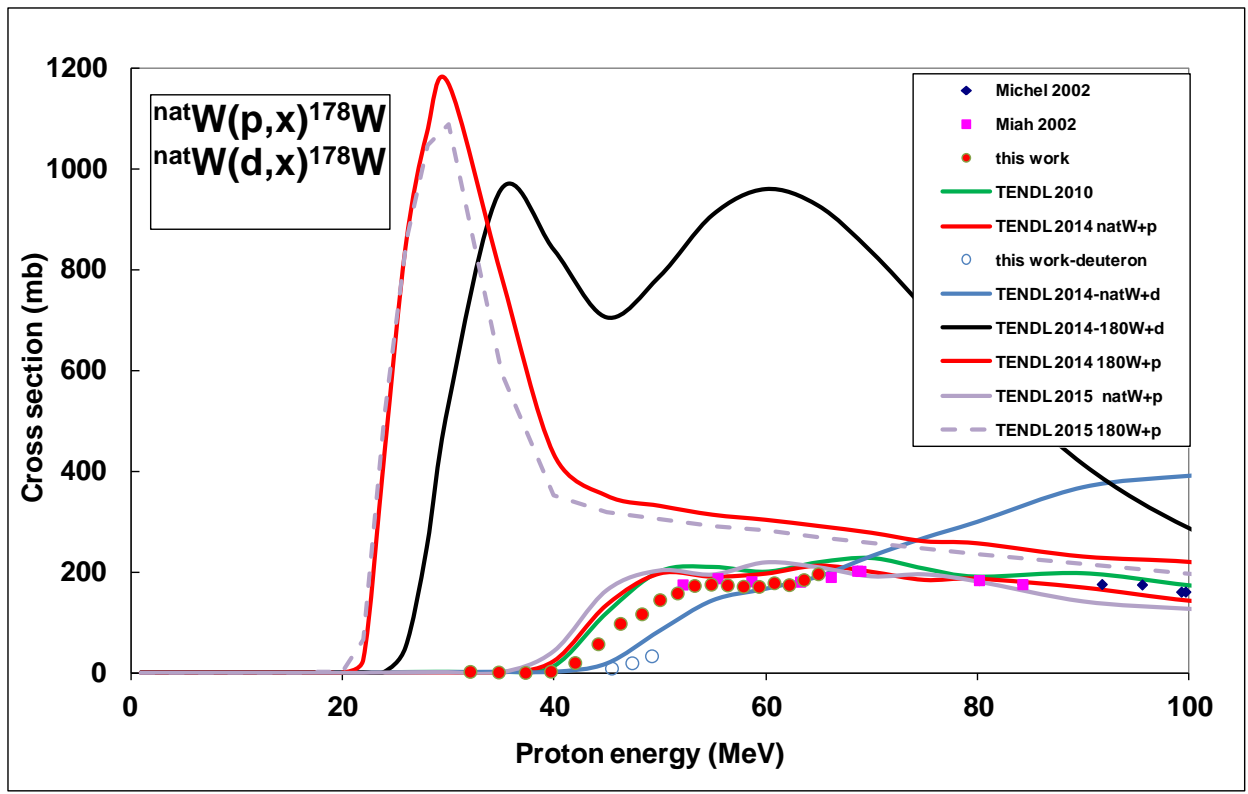

Fig. 8. Experimental excitation function for the $^{nat}$W(p,x)$^{178}$Re reaction and comparison with literature values and theoretical code calculations

## 4.7    Production of ¹⁸³Ta

The ¹⁸³Ta (5.1 d) can only be produced via ¹⁸⁴W (p,2p) and ¹⁸⁶W (p,2p2n) or ¹⁸⁶W (p,α) reactions and through the decay of ¹⁸³Hf (1.067 h).The low energy experimental data show large disagreements. It should be mentioned that the cross sections in the low energy region of (p,α) and (p,2pxn) reactions for this mass region are less than 10 mb, therefore all earlier experimental data at low energies seem to be very strange. Re-evaluating our data in [20], we found that the contribution of the overlapping gamma-lines of ¹⁸¹Re (19.9 h, 353.6 keV, 0.44 %) to the 353.99 keV ¹⁸³Ta gamma-line was not properly subtracted. It could be the reason for the too high [25] data. Our new data are in good agreement with the earlier data of [14] at high energy and with the magnitude of the theoretical prediction (Fig. 9).

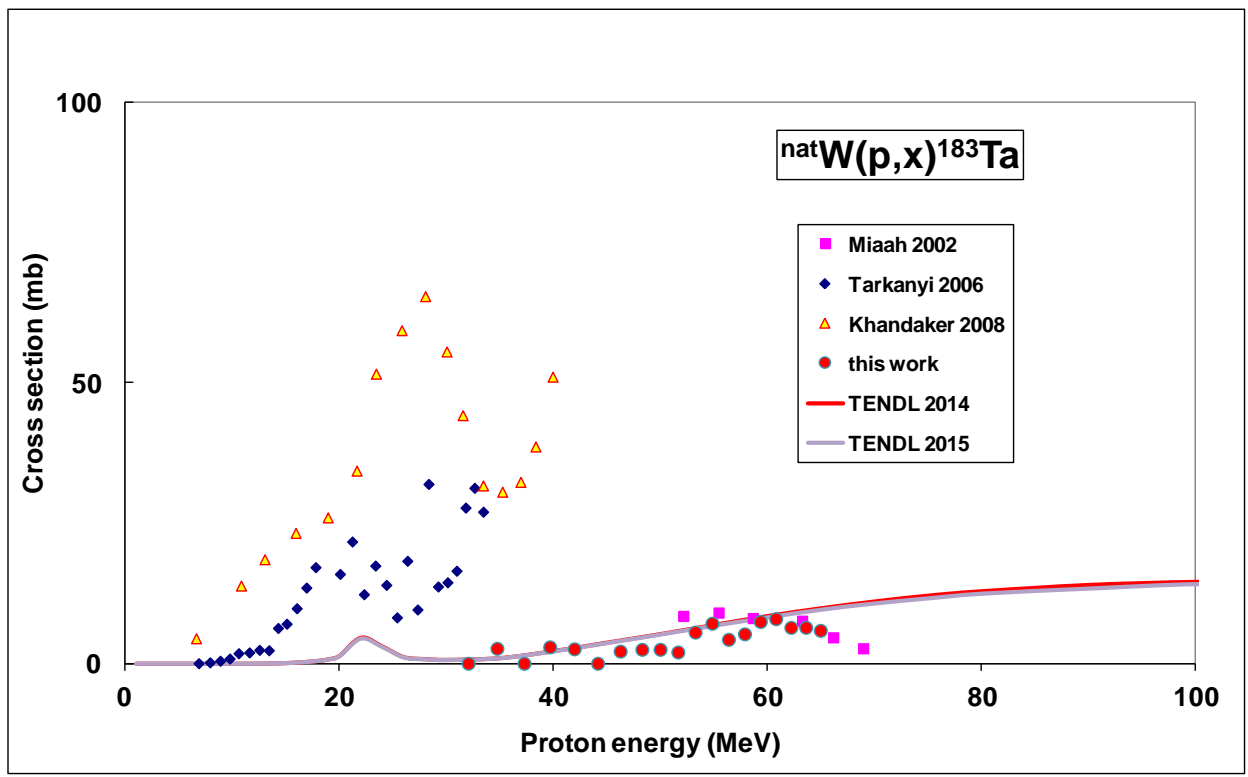

Fig. 9. Experimental excitation function for the $^{nat}$W(p,x)$^{183}$Ta reaction and comparison with literature values and theoretical code calculations

## 4.8    Production of $^{182}$Ta

The cumulative cross sections of $^{182}$Ta (114.74 d) ground state include the direct production and the production through the decay of $^{182m}$Ta (15.84 min, IT 100 %) and $^{182}$Hf (61.5 min, β- 55 %) isomeric states.  A good agreement with the high energy experiment of [14, 15] is seen (Fig. 10). The theory overestimates the experimental data.

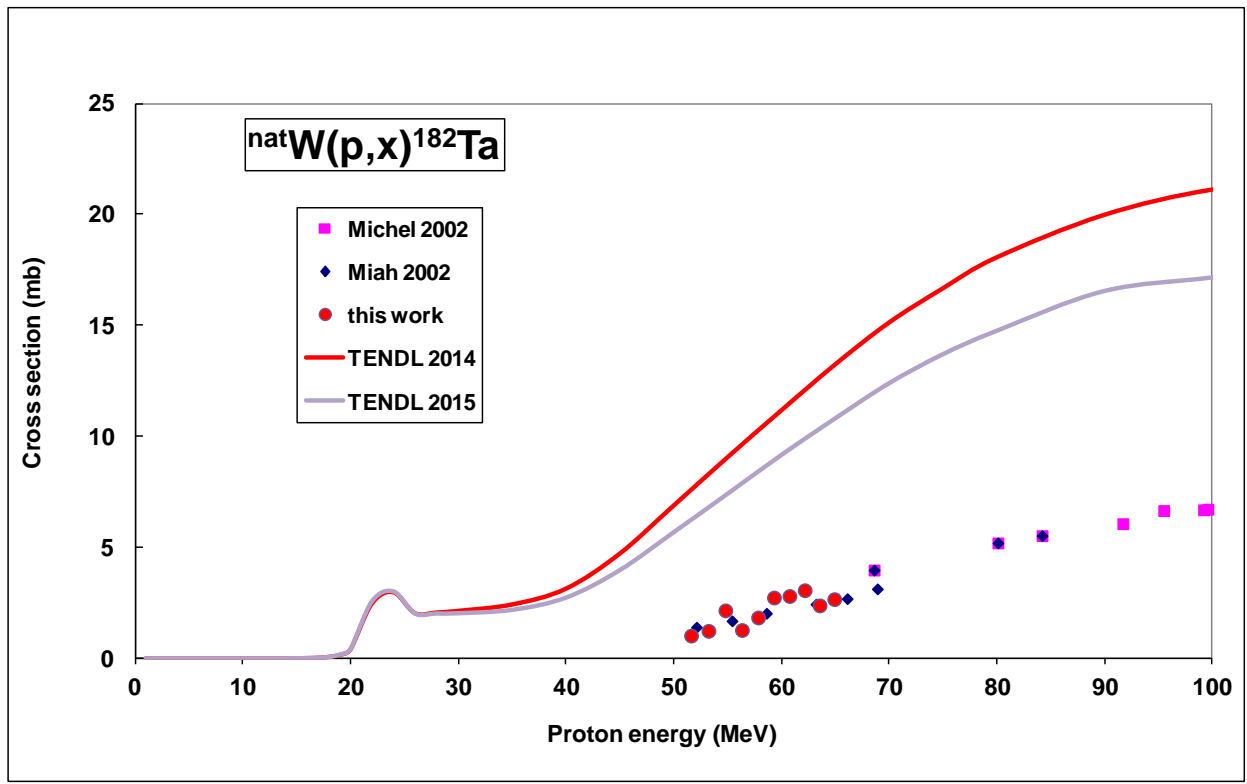

Fig. 10. Experimental excitation function for the $^{nat}$W(p,x)$^{182}$Ta reaction and comparison with literature values and theoretical code calculations.

### *4.9    Production of $^{180g}$Ta*

The   cross sections for formation of the $^{180}$Ta ground-state (8.154 h) are shown in Fig.11. It is produced directly and through the decay of $^{180m}$Hf (5.47 h, $\beta^-$: 0.31 %). The cross section values and uncertainties are influenced by the cooling time before measurement as the half-lives of $^{180g}$Ta and $^{180m}$Hf are similar.  Our experimental data are significantly higher than the predictions in the TENDL libraries.

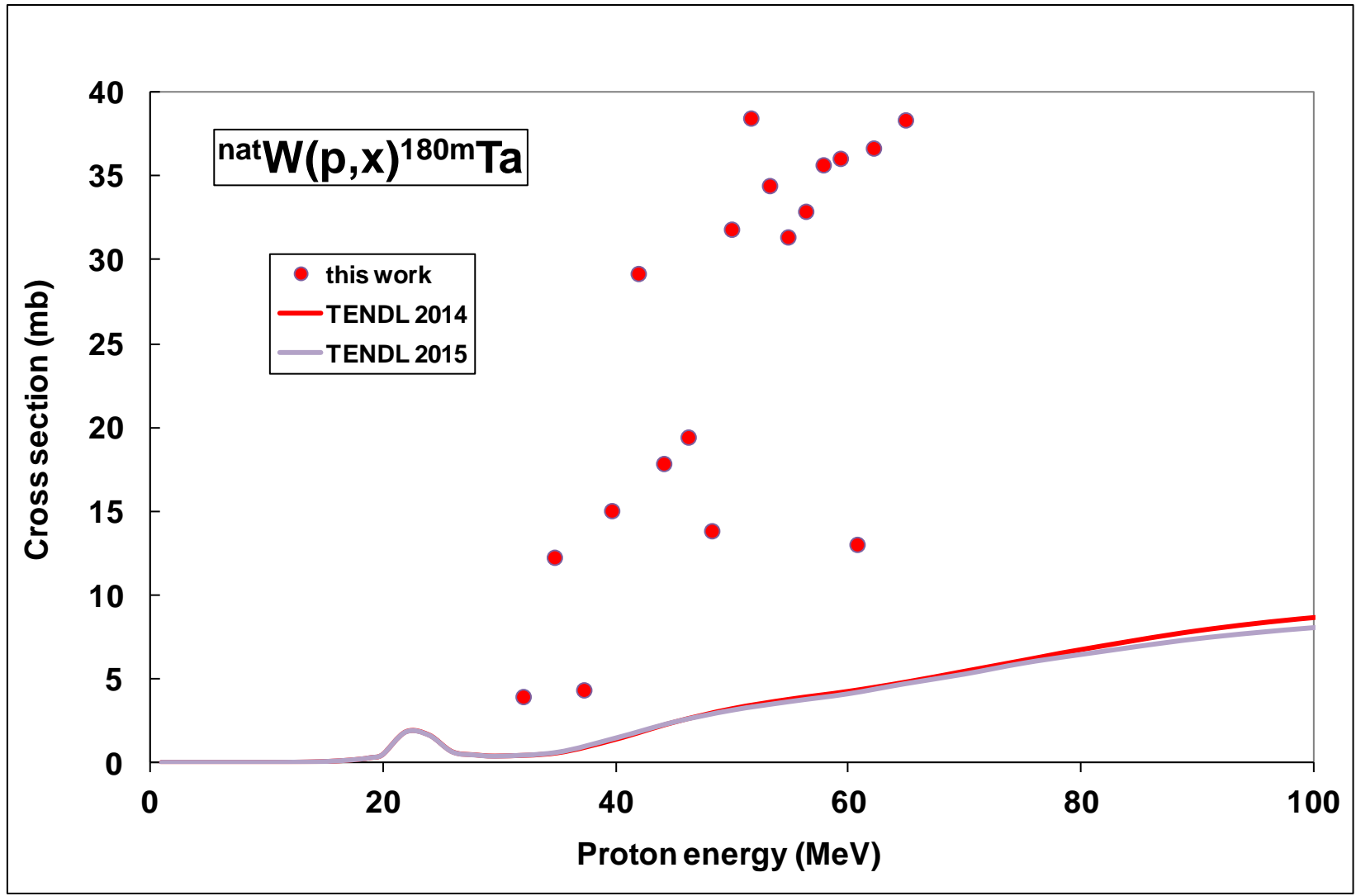

Fig. 11. Experimental excitation function for the $^{nat}W(p,x)^{180g}Ta$ reaction and comparison with literature values and theoretical code calculations

### *4.10 Production of $^{177}$Ta*

The radionuclide $^{177}$Ta (56.56 h) is produced directly and through the $^{177}$Re(14 min, ε: 100 %) → $^{177}$W ( 132 min, ε: 100 %) → $^{177}$Ta decay chain. There is a very significant disagreement between the experimental and theoretical data below 50 MeV (Fig. 12). Considering the systematics of the experimental magnitudes of the possible contributing reactions and their Q-values the reported literature data below 40 MeV are not realistic. As stated before the (p,2pxn) cross sections have maxima that are mostly below 10 mb. The reason for the too high reported cross sections is probably that the used gamma overlapped with a similar energy gamma-line emitted in the decay of another isotope, resulting in overestimation of the induced $^{177}$Ta activity.  In a re-evaluation of  the [20] study we found that  the used 112.9 keV line  contained contributions from the overlapping gamma-lines of $^{182}$Re (64.2 h, 111.07 keV, 0.209 %) and $^{184}$Re (35.4 d,111.2174 keV, 17.2 %) while for  the 208 keV  line the contributions of $^{182}$ Re( 64.2 h 208.26 keV, 0.62 %)  and $^{183}$Re (70.0 d , 208.8107 keV, 2.95 %) were not properly separated. It could be the reason for the discrepant results of [18].  Our new, well corrected data in the higher energy range agree with [14] and fit well with the predictions of TENDL.

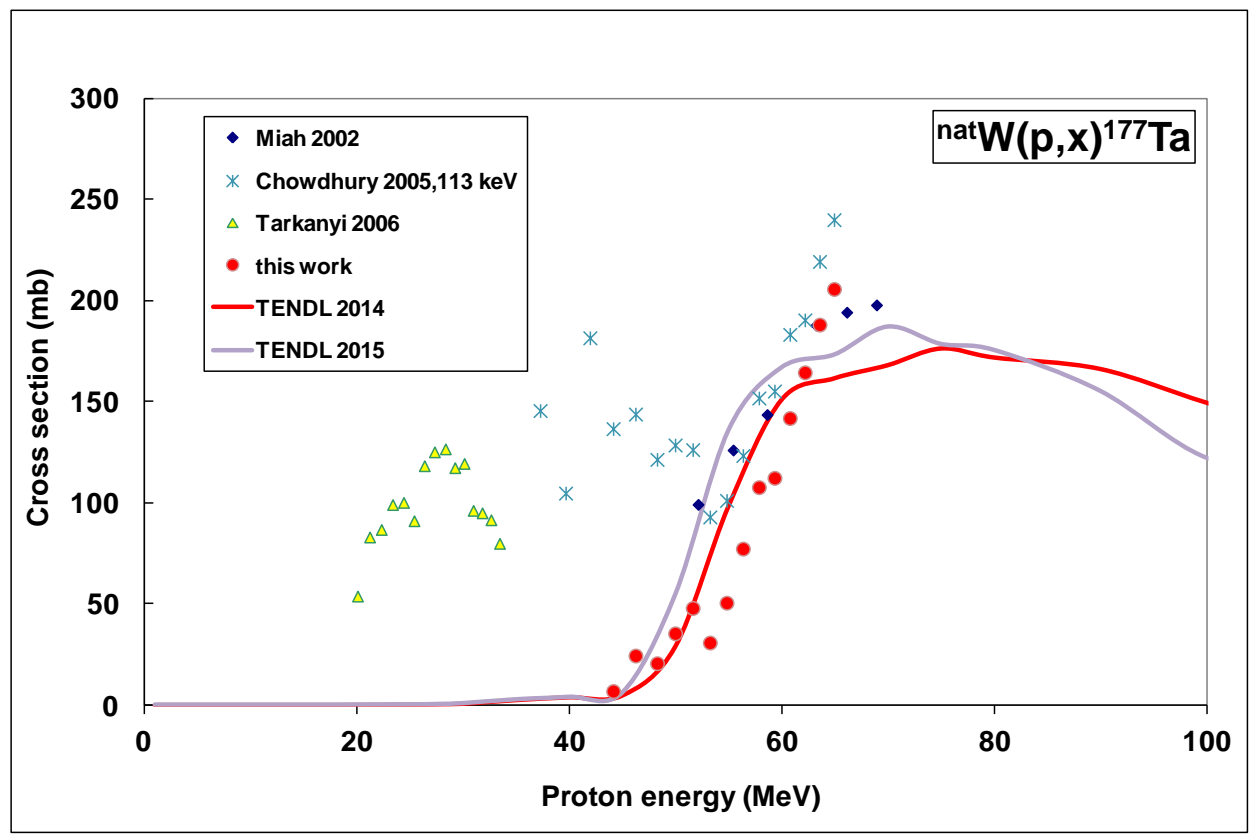

Fig. 12. Experimental excitation function for the $^{nat}$W(p,x)$^{177}$Ta reaction and comparison with literature values and theoretical code calculations

### 4.11 Production of $^{176}$Ta

The cross sections for $^{176}$Ta (8.09 h) are also cumulative as apart from reactions on all stable W isotopes also the $^{176}$Re (5.3 min, ε: 100 %) → $^{176}$W (2.5 h, ε: 100 %)→$^{176}$Ta decay chain contributes.  No evidence of emission of α-particle is seen. The agreement of the experimental data and the theoretical predictions is acceptable (Fig. 13).

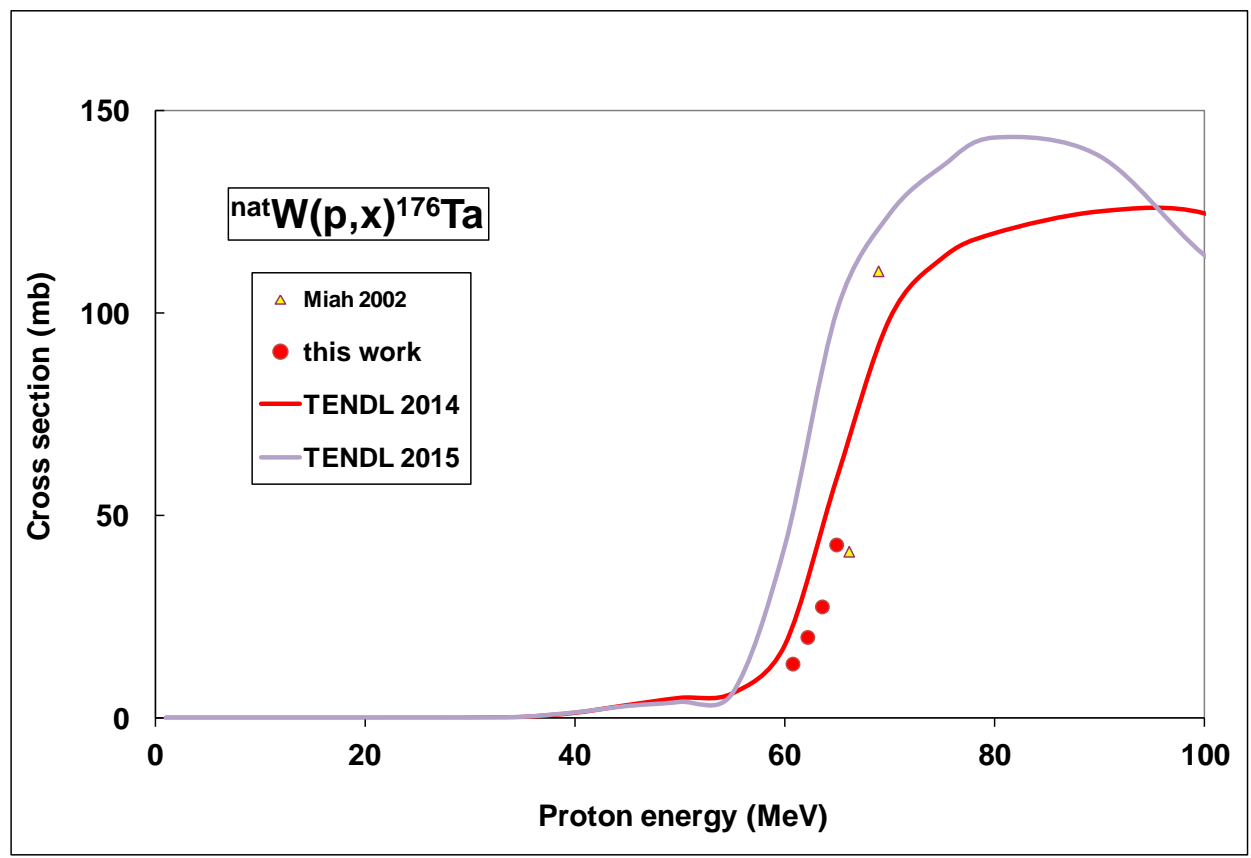

Fig. 13. Experimental excitation function for the natW(p,x)176Ta reaction and comparison with literature values and theoretical code calculations.

**4.12  Production of 175Ta**

The radionuclide $^{175}$Ta (10.5 h) is produced directly via (p,2pxn) reactions and through the $^{175}$Re (5.89 min, ε: 100 %) →$^{175}$W (35.2 min, ε: 100 %) →$^{75}$Ta decay chain.  As the threshold for the $^{182}$W(p,2p6n) reaction is more than 51 MeV (lowest mass W stable isotope with abundance of more than 1%) and as channels with emission of an α-particle have very low cross sections we could only observe a few points for cumulative production of $^{175}$Ta. The theoretical predictions are in good agreement with the experimental data (Fig. 14).

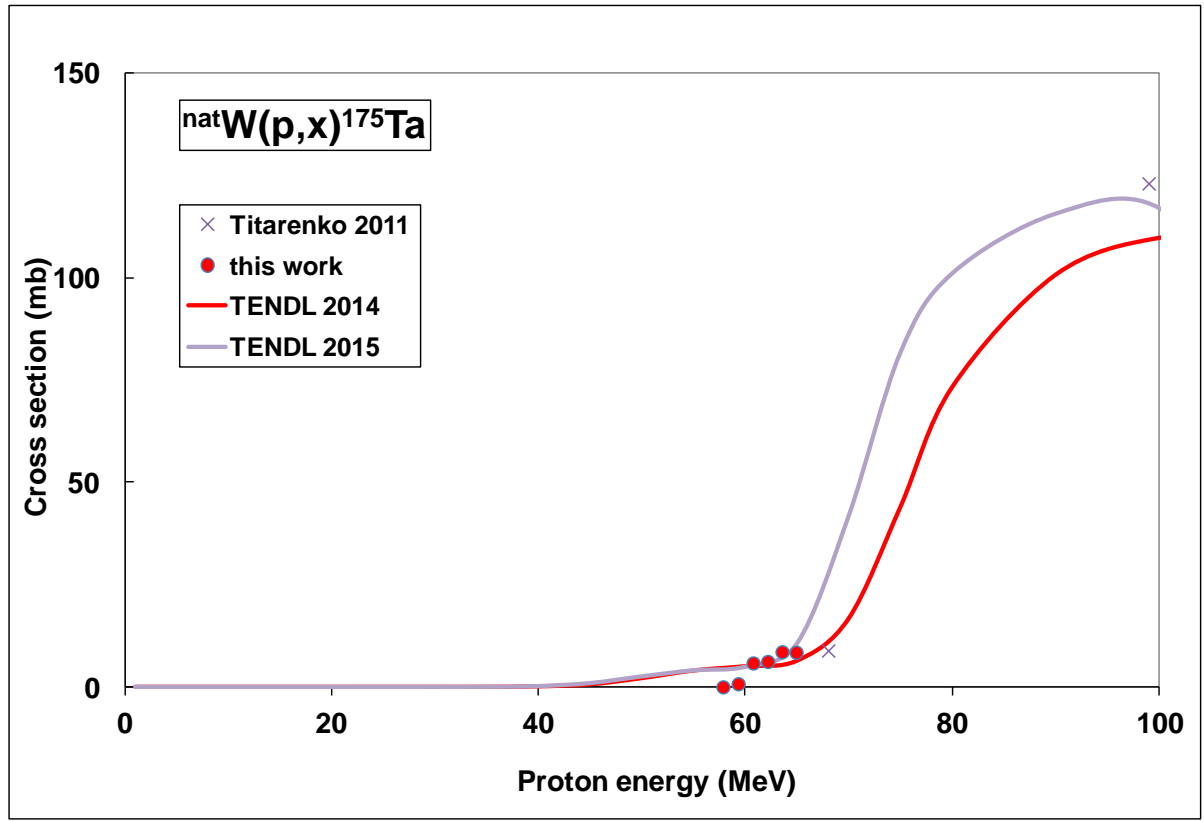

Fig. 14. Experimental excitation function for the $^{nat}$W(p,x)$^{175}$Ta reaction and comparison with literature values and theoretical code calculations

### 4.13   Production of $^{175}$Hf

The measured activation cross sections for $^{175}$Hf (70 d) are cumulative (Fig. 15). They include direct formation through (p,3pxn) reactions and through the $^{175}$Re (5.89 min, ε: 100 %) →$^{175}$W (35.2 min, ε: 100 %) →$^{175}$Ta (10.5 h) →$^{175}$Hf decay chain.  Both pathways have high thresholds and only cross sections at higher energy could be determined. A good agreement with the values of [14, 16] and with the predictions of TENDL are seen.

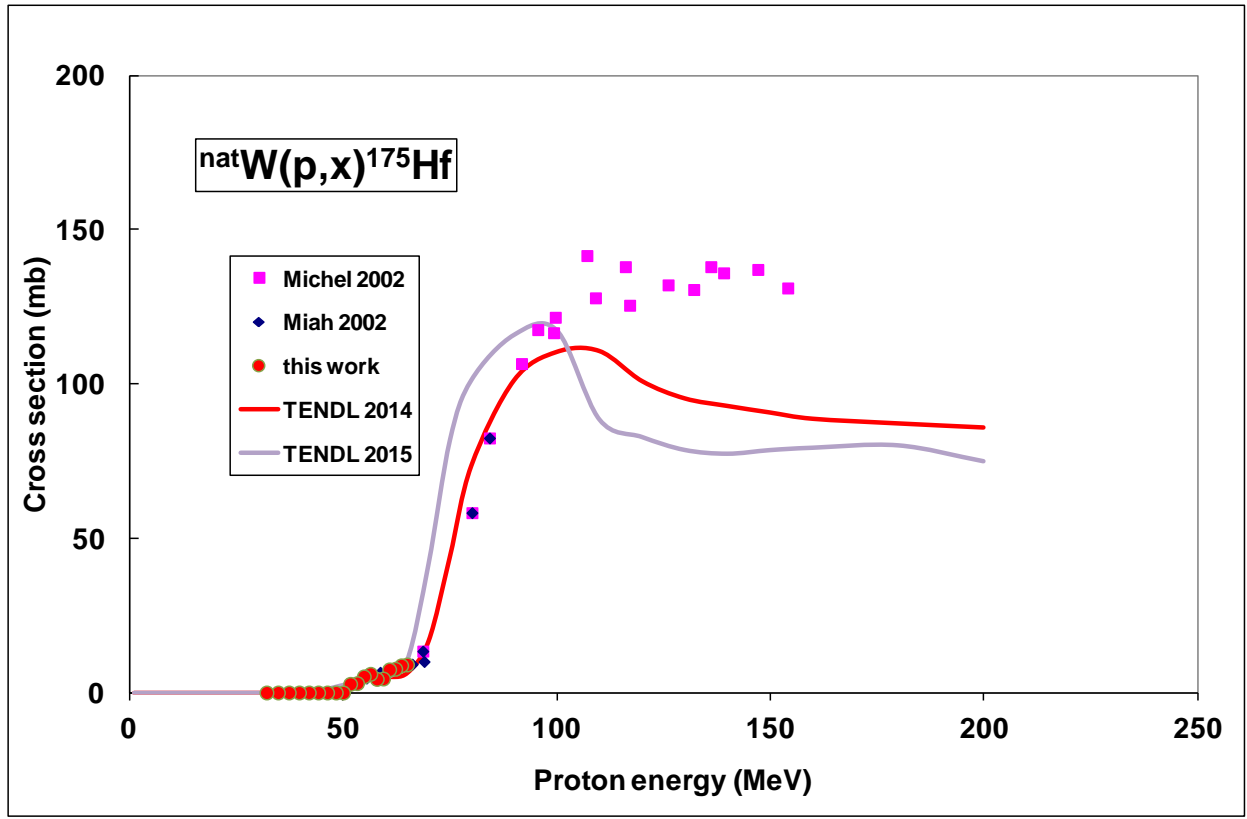

Fig. 15. Experimental excitation function for the $^{nat}$W(p,x)$^{175}$Hf reaction and comparison with literature values and theoretical code calculations

### 4.14 Production of $^{177}$Lu

The radionuclide $^{177}$Lu has two observable isomeric states: the higher laying $^{177m}$Lu (160.44 d, 23/2$^-$, β$^-$: 78.6 %, IT: 21.4 *8* %) metastable state and the shorter-lived ground-state $^{177g}$Lu (6.647 d). We could deduce cross section data for the production of the $^{177}$Lu ground state. It can be formed directly by high threshold (p,4pxn) reactions and from the β$^-$ decay of $^{177}$Yb (1.911 h, β⁻: 100 %). As no signals were observed from the decay of $^{177m}$Lu the possible contribution to ground state formation was neglected. Our data differ significantly from earlier results of [14]. The theory predictions are significantly lower than both experimental data sets (Fig. 16).

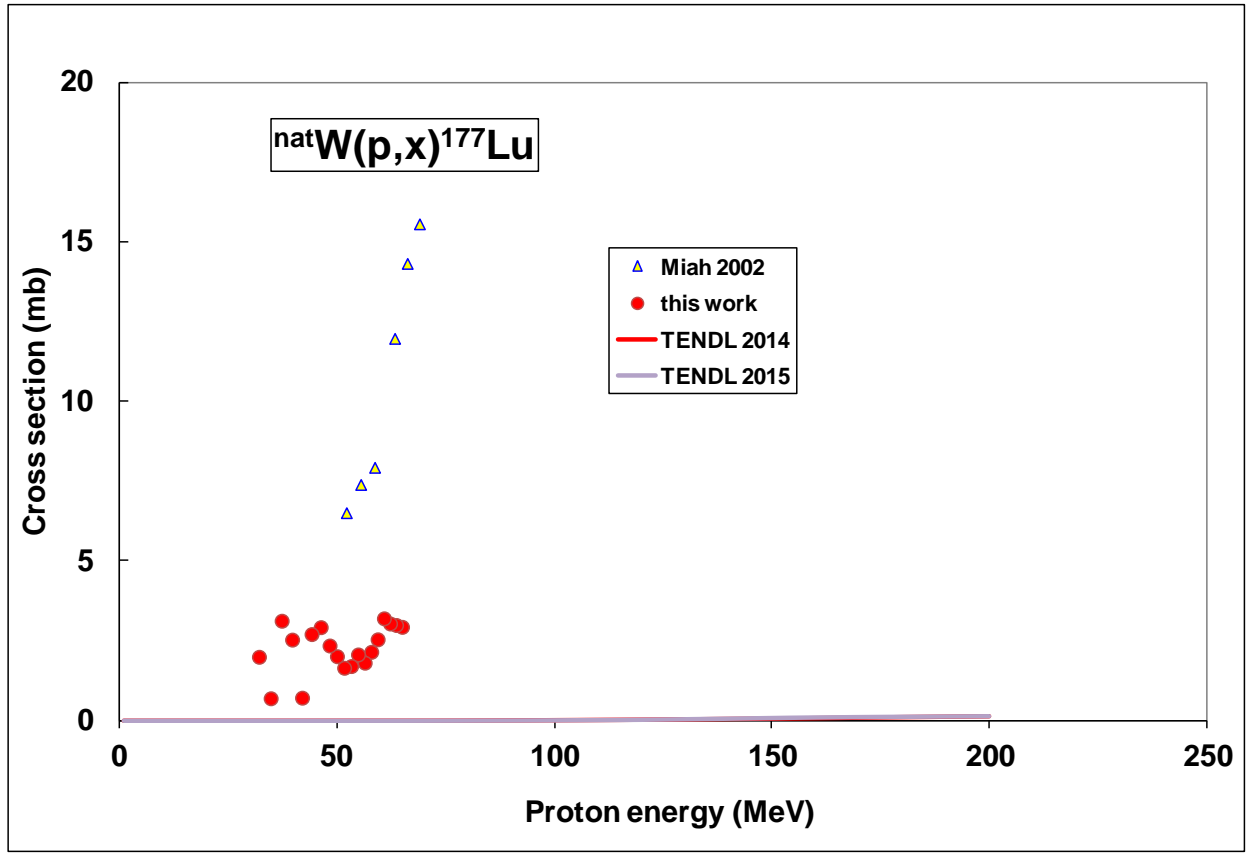

Fig. 16. Experimental excitation function for the $^{nat}$W(p,x)$^{177}$Lu reaction and comparison with literature values and theoretical code calculations

## 5. Integral yields

Integrated yields for a given incident energy down to the reaction threshold were calculated as a function of the incident energy from fitted curves to our experimental cross section data. The results for physical yields [39, 40] are presented in Figs. 17 and 18 in comparison with the experimental yield data of [7, 26, 41-43] …..

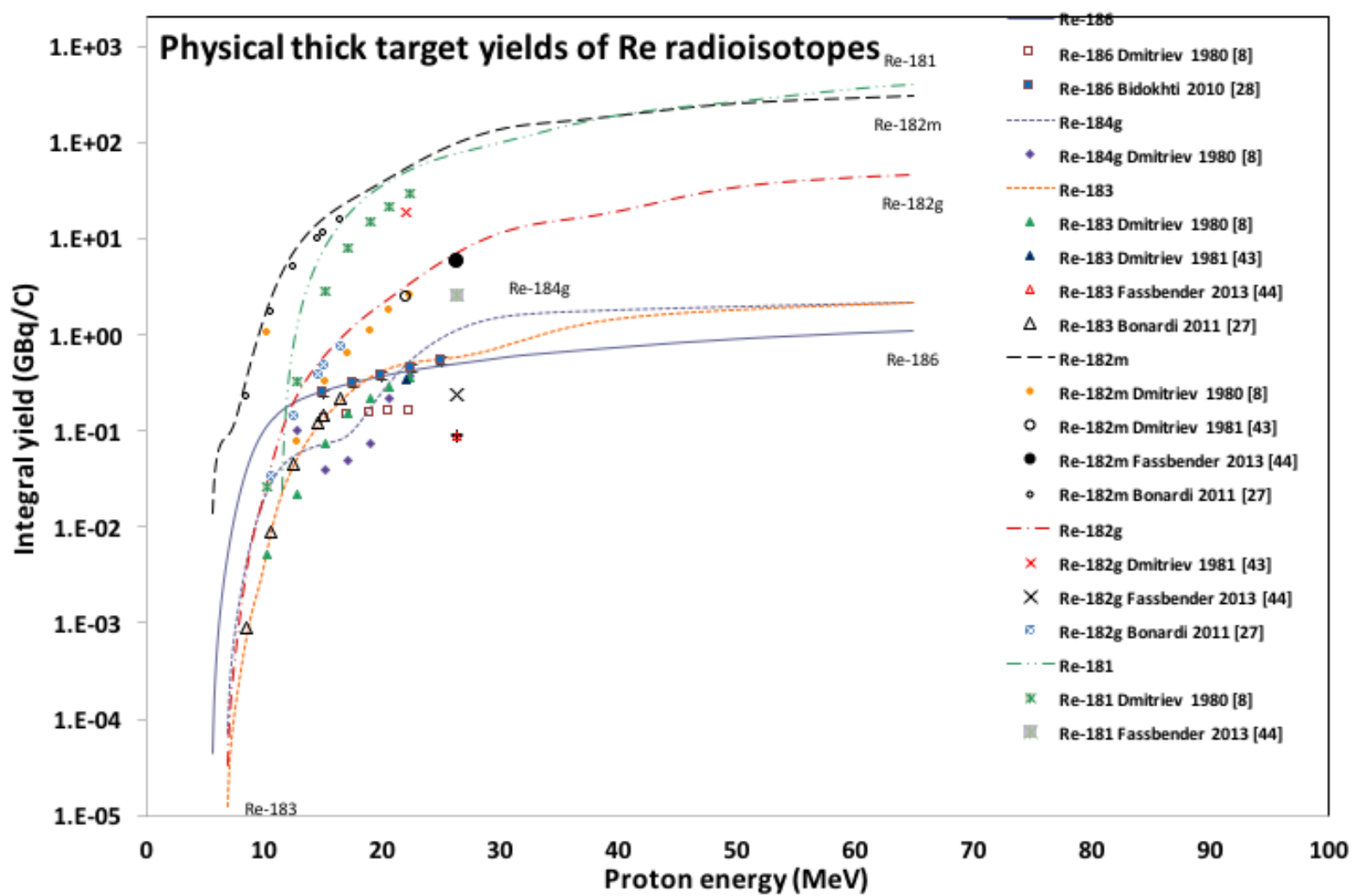

Fig. 17. Integral yields for production of Re radioisotopes deduced from the excitation functions in comparison with the literature experimental integral yields.

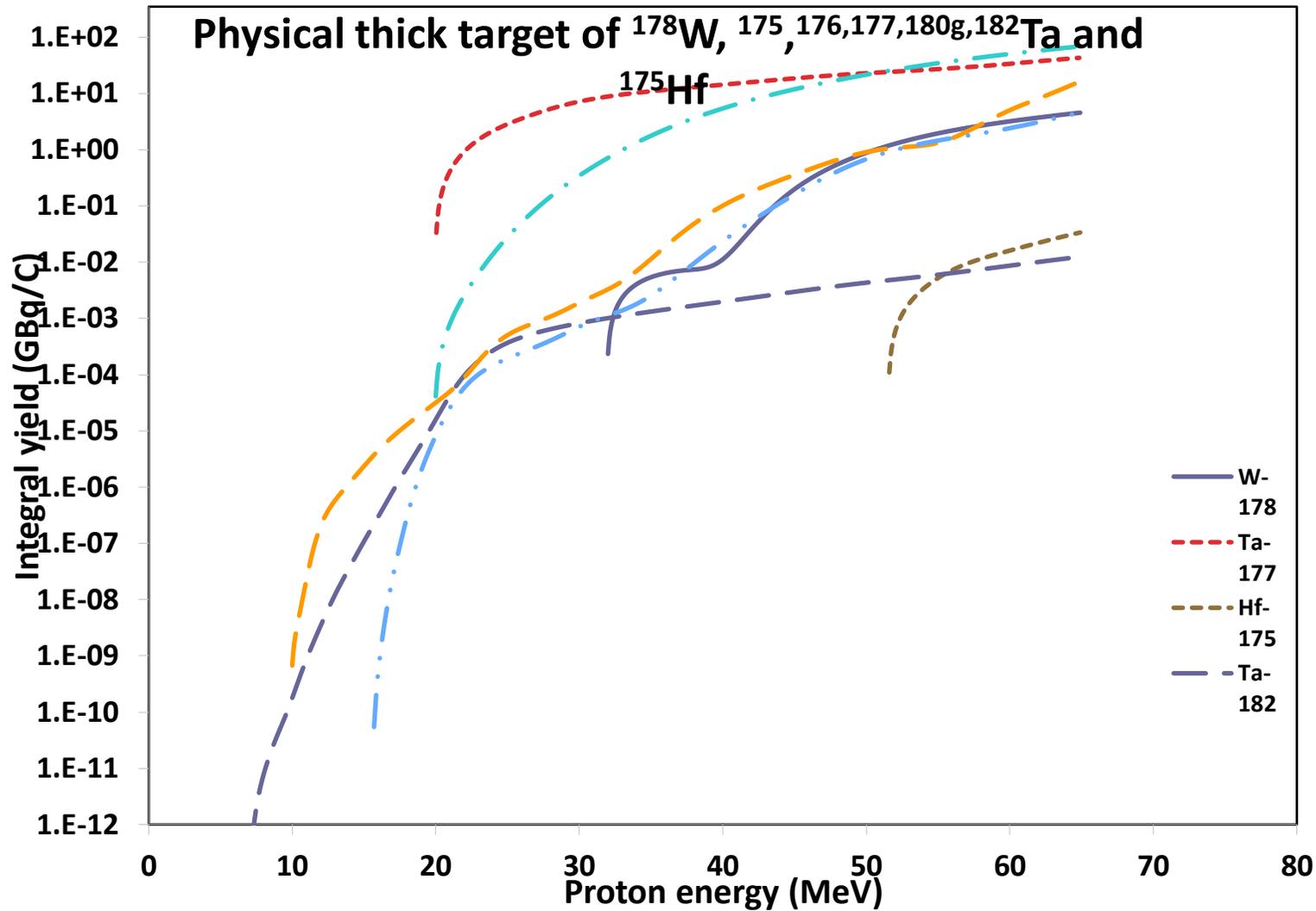

Fig. 18. Integral yields for production of [178]W, [177]Ta and [175]Hf deduced from the excitation functions in comparison with the literature experimental integral yields.

## 6. Production routes of medically relevant radioisotopes

Among the studied reaction products $^{186}$Re, $^{178}$W/$^{178}$Ta and $^{181}$W have medically relevant applications and their different possible production routes will be qualitatively discussed here.

### 6.1 Production of $^{186}$Re

Different routes for production of $^{186}$Re were measured recently by many authors, and also by us [1, 20]. In this work only high energy part of $^{186}$W(p,n) reaction was measured having small practical importance for $^{186}$Re production. The recently investigated $^{186}$W(p,n)$^{186}$Re reaction together with the $^{186}$W(d,2n)$^{186}$Re reaction can be considered for the production of the therapeutic radioisotope $^{186}$Re in no-carrier-added form. The $^{186}$Re yield is significantly higher in deuteron induced reaction compared to the widely used proton irradiation, but it requires higher energy machines and stable $^{187}$Re will be coproduced, lowering specific activity.

### 6.2 Production of $^{181}$W

Many charged particle induced routes were studied for production of $^{181}$W: it is produced by direct formation via $^{181}$Ta(p,n), $^{181}$Ta(d,2n) and $^{180}$Hf($\alpha$,2n) reactions and indirectly via the $^{182}$W(p,2n)$^{181}$Re→$^{181}$W, $^{nat}$W(p,xn)$^{181}$Re→$^{181}$W routes. Most of these routes were also investigated by us [2, 20, 44, 45]. Out of the direct routes the (d,2n) reaction is the most productive. The used tantalum target is practically monoisotopic (99.988%). The indirect route relying on $^{182}$W (p,2n) or $^{nat}$W(p,xn) is even more productive, but it requires enriched targets or in case of natural tungsten the specific activity of the $^{181}$W is lower.

### 6.3 Production of $^{178}$W/$^{178m}$Ta generator

The production routes for $^{178}$W were reviewed recently by us [2]. The following reactions were included in the review:

- Proton and deuteron induced spallation evaporation on heavy mass targets

- Proton induced route via the $^{181}$Ta(p,4n) reaction
- Deuteron (50 MeV) route via the $^{181}$Ta(d,5n) reaction
- α-particle (50-30 MeV incident energy) route via $^{nat}$Hf(α,x) and $^{176,177,178}$Hf(α,xn) reactions
- $^{3}$He induced route via $^{nat}$Hf($^{3}$He,x) and $^{176,177,178}$Hf($^{3}$He,xn) reactions.

Spallation requires high energy accelerators. The proton induced reaction on tantalum requires incident energy up to 70 MeV for high yield. The optimum energy range for the deuteron route (70- 35 MeV) is currently not available for accelerators for routine production. Alpha induced reactions on hafnium could be implemented at low energy accelerators but the production yield will be lower (compared to the proton and deuteron routes). The $^{3}$He-induced reactions are not competitive from all points of view.

The light ion activation routes of the $^{178}$W for production of the medically related $^{178}$W/$^{178}$Ta generator showed that in the medium energy range the $^{181}$Ta(p,4n), at low energies the $^{176}$Hf(α,2n) reactions are the favorite routes.

In principle this review can be completed with the recently investigated $^{nat}$W(p,x)$^{178}$W and $^{nat}$W(d,x)$^{178}$W reactions (see Fig. 8). The main difference is that the previously discussed production routes are not carrier added for production of the $^{178}$W, in opposition to using W targets.

In case of proton bombardment of W targets the longer-lived W-activation products are $^{178}$W (21.6 d), $^{181}$W (121.2 d), $^{185}$W (75.1 d). The latter two are decaying to stable $^{181}$Ta and $^{185}$Re, therefore in the eluted product only $^{178}$Ta and the stable $^{181}$Ta will be present. Considering the production yield of the $^{nat}$W(p,x)$^{178}$W the effective threshold is around 40 MeV, but the cross sections are high (200 mb) up to 100 MeV.

To get some experimental data $^{nat}$W(d,x)$^{178}$W for this work  we have re-evaluated the spectra from our earlier irradiation in 1998 (see Fig. 19). We have reported all other experimental results in [1], but the evaluation of $^{178}$W was missed, due to the preliminary estimated high reaction threshold (we refer to experimental details to [1]. According to Fig. 18, the effective reaction threshold for $^{nat}$W(d,x)$^{178}$W reaction is around 45 MeV. The predicted cross section around 60 MeV also reaches 200 mb, and around 100 MeV the maximum is near 400 mb.

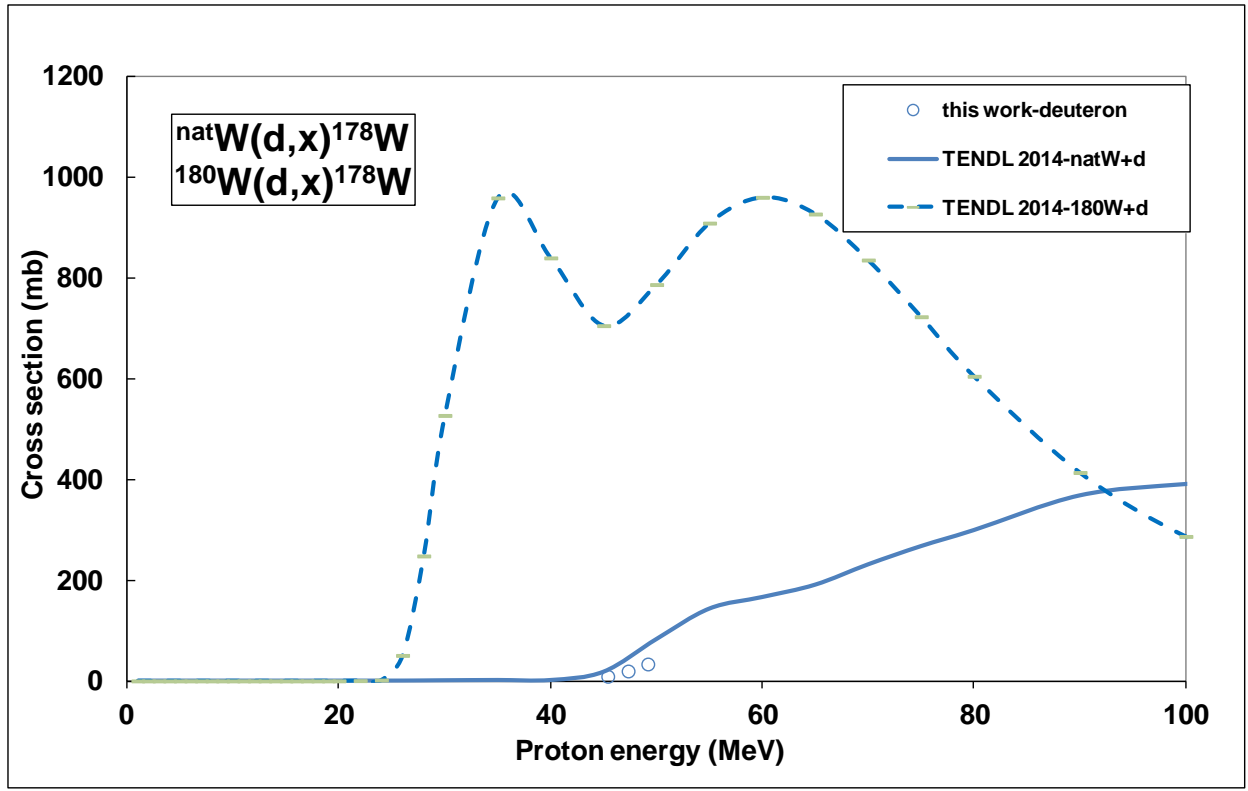

Fig. 19. Experimental excitation function for the $^{nat}$W(d,x)$^{178}$Re reaction and comparison with literature values and theoretical code calculations

Therefore, in both proton and deuteron induced reactions on $^{nat}$W, the yield is high, it requires high energy machines and the product is not carrier free.

Fig. 7 and Fig. 18 show that by using highly enriched $^{180}$W target, the production of $^{178}$W is possible with high yields already at lower energy machines, especially in case of $^{180}$W(p,p2n)$^{178}$W.  In case of protons no longer-lived $^{181}$W and $^{185}$W byproduct will be formed and the eluted product is carrier free. In case of deuteron irradiation no $^{185}$W is formed, but $^{181}$W will be still present via (d,p).

A disadvantage of this route is the very low isotopic abundance of $^{180}$W (0.13 %) in $^{nat}$W, i.e. the price of the highly enriched $^{180}$W target will be a limiting factor.

## 7. Summary

We present activation cross section data for proton induced reactions on natural tungsten for production of for $^{nat}$W(p,xn)$^{186,184,183,182m,182g,181}$Re, $^{na}$W(p,x)$^{178}$W, $^{nat}$W(p,x)$^{183,182,180m,177,176,175}$Ta, $^{175}$Hf and $^{177}$Lu  in the 32-65 MeV energy range and additionally  a few data for $^{nat}$W(d,x)$^{178}$W near 50 MeV. For some reactions the data are measured for the first time in this energy region. The excitation functions are correlated to monitor reactions simultaneously re-measured over the whole energy range. In average our new experimental data are in good agreement with the earlier reported medium energy experimental data. Comparison with TENDL-2014 and 2015 theoretical predictions shows acceptable agreement. Based on a detailed evaluation of possible contributing reactions and a comparison with the TENDL predictions a necessity of correction of some earlier reported data in the  low energy range was found. A short qualitative review of production routes of medically relevant $^{186}$Re,$^{181}$W,$^{178}$W/$^{178m}$Ta radioisotopes are given with emphasis of practical use of activation cross section data of proton and deuteron induced reactions on tungsten.

*Acknowledgements*

This work was done in the frame of MTA-FWO (Vlaanderen) research projects. The authors acknowledge the support of research projects and of their respective institutions in providing the materials and the facilities for this work.

**FIGURE CAPTIONS**

Fig. 1. Experimental excitation function for the $^{nat}$W(p,x)$^{186}$Re reaction and comparison with literature values and theoretical code calculations

Fig. 2. Experimental excitation function for the $^{nat}$W(p,x)$^{184m}$Re reaction and comparison with literature values and theoretical code calculations

Fig. 3. Experimental excitation function for the $^{nat}$W(p,x)$^{184g}$Re reaction and comparison with literature values and theoretical code calculations

Fig. 4. Experimental excitation function for the $^{nat}$W(p,x)$^{183}$Re reaction and comparison with literature values and theoretical code calculations

Fig. 5. Experimental excitation function for the $^{nat}$W(p,x)$^{182m}$Re reaction and comparison with literature values and theoretical code calculations

Fig. 6. Experimental excitation function for the $^{nat}$W(p,x)$^{182g}$Re reaction and comparison with literature values and theoretical code calculations

Fig. 7. Experimental excitation function for the $^{nat}$W(p,x)$^{181}$Re reaction and comparison with literature values and theoretical code calculations

Fig. 8. Experimental excitation function for the $^{nat}$W(p,x)$^{178}$Re reaction and comparison with literature values and theoretical code calculations

Fig. 9. Experimental excitation function for the $^{nat}$W(p,x)$^{183}$Ta reaction and comparison with literature values and theoretical code calculations

Fig. 10. Experimental excitation function for the $^{nat}W(p,x)^{182}Ta$ reaction and comparison with literature values and theoretical code calculations

Fig. 11. Experimental excitation function for the $^{nat}W(p,x)^{180g}Ta$ reaction and comparison theoretical code calculations

Fig. 12. Experimental excitation function for the $^{nat}W(p,x)^{177}Ta$ reaction and comparison with literature values and theoretical code calculations

Fig. 13. Experimental excitation function for the $^{nat}W(p,x)^{176}Ta$ reaction and comparison with literature values and theoretical code calculations

Fig. 14. Experimental excitation function for the $^{nat}W(p,x)^{175}Ta$ reaction and comparison with literature values and theoretical code calculations

Fig. 15. Experimental excitation function for the $^{nat}W(p,x)^{175}Hf$ reaction and comparison with literature values and theoretical code calculations

Fig. 16. Experimental excitation function for the $^{nat}W(p,x)^{177}Lu$ reaction and comparison with literature values and theoretical code calculations

Fig. 17. Integral yields for production of Re radioiotopes deduced from the excitation functions in comparison with the literature experimental integral yields.

Fig. 18. Integral yields for production of $^{178}W$, $^{177}Ta$ and $^{175}Hf$ deduced from the excitation functions in comparison with the literature experimental integral yields.

Fig. 19. Experimental excitation function for the $^{nat}$W(d,x)$^{178}$Re reaction and comparison with literature values and theoretical code calculations

**Table 2** Experimental cross sections of $^{nat}W(p,xn)^{186,184,183,\ 182m,182g,181}Re$

| E±ΔE (MeV) | | σ± Δσ(mb) | | | | | | | | | | |
|---|---|---|---|---|---|---|---|---|---|---|---|---|
| | | 186Re | | 184Re | | 183Re | | 182mRe | | 182gRe | | 181Re | |
| 64.9 | 0.2 | 1.29 | 0.16 | 13.61 | 1.56 | 43.89 | 4.95 | 69.05 | 17.93 | 48.90 | 5.83 | 183.75 | 20.66 |
| 63.5 | 0.3 | 1.37 | 0.17 | 13.64 | 1.57 | 45.39 | 5.12 | 68.53 | 19.35 | 44.59 | 5.40 | 168.94 | 19.01 |
| 62.2 | 0.3 | 1.36 | 0.17 | 14.27 | 1.64 | 44.40 | 5.01 | 45.72 | 14.38 | 37.00 | 4.55 | 217.36 | 24.44 |
| 60.7 | 0.4 | 1.31 | 0.17 | 15.26 | 1.76 | 45.55 | 5.15 | 64.29 | 15.65 | 59.06 | 7.12 | 246.17 | 27.66 |
| 59.3 | 0.5 | 1.46 | 0.18 | 14.96 | 1.70 | 47.25 | 5.32 | 50.40 | 17.86 | 61.30 | 7.24 | 283.93 | 31.92 |
| 57.8 | 0.6 | 1.83 | 0.22 | 14.94 | 1.72 | 50.73 | 5.72 | 48.86 | 16.87 | 62.95 | 7.39 | 258.78 | 29.07 |
| 56.3 | 0.6 | 1.54 | 0.19 | 16.00 | 1.82 | 50.63 | 5.70 | 61.35 | 18.11 | 65.77 | 7.72 | 283.09 | 31.81 |
| 54.8 | 0.7 | 1.65 | 0.20 | 16.91 | 1.94 | 53.91 | 6.08 | 76.85 | 21.00 | 72.23 | 8.44 | 278.07 | 31.25 |
| 53.2 | 0.8 | 1.66 | 0.23 | 17.46 | 2.03 | 61.56 | 6.96 | 75.22 | 22.08 | 88.71 | 10.30 | 270.44 | 30.40 |
| 51.6 | 0.9 | 1.86 | 0.24 | 17.60 | 2.02 | 59.22 | 6.67 | 81.78 | 19.36 | 111.81 | 12.75 | 264.24 | 29.69 |
| 49.9 | 0.9 | 2.12 | 0.27 | 18.05 | 2.09 | 63.86 | 7.20 | 118.32 | 26.36 | 132.98 | 15.16 | 245.01 | 27.53 |
| 48.2 | 1.0 | 2.27 | 0.27 | 17.94 | 2.05 | 68.32 | 7.69 | 112.80 | 23.26 | 143.93 | 16.49 | 212.59 | 23.90 |
| 46.2 | 1.1 | 2.04 | 0.26 | 21.01 | 2.41 | 84.60 | 9.53 | 155.73 | 31.20 | 170.50 | 19.41 | 205.20 | 23.07 |
| 44.1 | 1.2 | 2.75 | 0.34 | 21.19 | 2.43 | 95.83 | 10.79 | 182.10 | 35.80 | 185.28 | 21.04 | 238.80 | 26.84 |
| 41.9 | 1.3 | 2.50 | 0.30 | 24.48 | 2.78 | 118.36 | 13.31 | 136.08 | 32.04 | 145.39 | 16.57 | 281.57 | 31.64 |
| 39.6 | 1.5 | 2.60 | 0.32 | 27.09 | 3.09 | 160.58 | 18.05 | 181.50 | 30.31 | 126.61 | 14.41 | 345.29 | 38.78 |
| 37.2 | 1.6 | 2.59 | 0.31 | 30.52 | 3.45 | 221.70 | 24.90 | 144.71 | 28.95 | 83.28 | 9.70 | 388.97 | 43.70 |
| 34.7 | 1.7 | 2.94 | 0.34 | 36.99 | 4.16 | 266.90 | 29.96 | 98.56 | 23.60 | 70.95 | 8.38 | 386.55 | 43.43 |
| 32.0 | 1.8 | 2.79 | 0.32 | 50.41 | 5.67 | 257.43 | 28.90 | 136.11 | 15.83 | 102.19 | 11.53 | 267.53 | 30.04 |
| | | | | | | | | | | | | | |

**Table 3** Experimental cross sections of, $^{nat}W(p,x)^{183,182,\ 180m,\ 177,176,175}Ta$

| E±ΔE (MeV) | σ± Δσ(mb) |
|---|---|

| | | 183Ta | | 182Ta | | 180mTa | | 177Ta | | 176Ta | | 175Ta | |
|---|---|---|---|---|---|---|---|---|---|---|---|---|---|
| 64.9 | 0.2 | 5.87 | 0.70 | 2.65 | 0.62 | 38.36 | 4.77 | 205.70 | 48.90 | 42.82 | 5.57 | 8.46 | 1.52 |
| 63.5 | 0.3 | 6.41 | 0.78 | 2.36 | 0.54 | 43.64 | 5.61 | 188.08 | 44.59 | 27.53 | 3.87 | 8.57 | 1.50 |
| 62.2 | 0.3 | 6.40 | 0.80 | 3.05 | 0.70 | 36.68 | 4.72 | 164.49 | 37.00 | 19.99 | 2.80 | 6.20 | 1.14 |
| 60.7 | 0.4 | 7.95 | 0.97 | 2.79 | 0.72 | 13.04 | 1.72 | 141.90 | 59.06 | 13.40 | 1.88 | 5.84 | 0.71 |
| 59.3 | 0.5 | 7.45 | 0.87 | 2.72 | 0.71 | 36.07 | 4.91 | 112.33 | 61.30 | | | 2.76 | 0.50 |
| 57.8 | 0.6 | 5.24 | 0.62 | 1.82 | 0.42 | 35.68 | 4.56 | 107.72 | 62.95 | | | | |
| 56.3 | 0.6 | 4.27 | 0.49 | 1.25 | 0.35 | 32.91 | 4.43 | 77.34 | 65.77 | | | | |
| 54.8 | 0.7 | 7.14 | 0.84 | 2.14 | 0.50 | 31.38 | 4.22 | 50.62 | 72.23 | | | | |
| 53.2 | 0.8 | 5.55 | 0.74 | 1.22 | 0.14 | 34.44 | 4.85 | 30.92 | 88.71 | | | | |
| 51.6 | 0.9 | 2.01 | 0.27 | 1.00 | 0.11 | 38.47 | 7.11 | 48.01 | 111.81 | | | | |
| 49.9 | 0.9 | 2.48 | 0.32 | | | 31.84 | 4.24 | 35.47 | 132.98 | | | | |
| 48.2 | 1.0 | 2.48 | 0.30 | | | 13.85 | 2.06 | 20.71 | 143.93 | | | | |
| 46.2 | 1.1 | 2.16 | 0.28 | | | 19.44 | 3.46 | 24.50 | 170.50 | | | | |
| 44.1 | 1.2 | | | | | 17.86 | 3.44 | 6.95 | 185.28 | | | | |
| 41.9 | 1.3 | 2.57 | 0.33 | | | 29.20 | 6.62 | | | | | | |
| 39.6 | 1.5 | 2.92 | 0.37 | | | 15.05 | 3.66 | | | | | | |
| 37.2 | 1.6 | | | | | 4.36 | 3.17 | | | | | | |
| 34.7 | 1.7 | 2.66 | 0.30 | | | 12.27 | 4.12 | | | | | | |
| 32.0 | 1.8 | | | | | 3.96 | 0.52 | | | | | | |
| | | | | | | | | | | | | | |
| | | | | | | | | | | | | | |

**Table 4** Experimental cross sections of $^{nat}W(p,x)^{178}W$, $^{nat}W(p,x)$ $^{175}Hf, ^{177}Lu$

| E±ΔE (MeV) | | σ± Δσ(mb) | | | | | | | | | | | |
|---|---|---|---|---|---|---|---|---|---|---|---|---|---|
| | | 178W | | 175Hf | | 177Lu | | | | | | | |
| 64.9 | 0.2 | 196.34 | 22.06 | 9.08 | 1.03 | 2.93 | 0.33 | | | | | | |
| 63.5 | 0.3 | 184.73 | 20.76 | 8.80 | 1.00 | 2.99 | 0.34 | | | | | | |
| 62.2 | 0.3 | 174.78 | 19.65 | 7.69 | 0.88 | 3.03 | 0.34 | | | | | | |
| 60.7 | 0.4 | 178.50 | 20.07 | 7.47 | 0.88 | 3.19 | 0.36 | | | | | | |
| 59.3 | 0.5 | 171.28 | 19.24 | 4.53 | 0.51 | 2.54 | 0.29 | | | | | | |
| 57.8 | 0.6 | 171.57 | 19.29 | 4.40 | 0.51 | 2.15 | 0.24 | | | | | | |
| 56.3 | 0.6 | 174.15 | 19.57 | 6.05 | 0.69 | 1.80 | 0.20 | | | | | | |
| 54.8 | 0.7 | 175.64 | 19.74 | 5.21 | 0.66 | 2.06 | 0.23 | | | | | | |
| 53.2 | 0.8 | 173.14 | 19.49 | 3.04 | 0.44 | 1.70 | 0.20 | | | | | | |
| 51.6 | 0.9 | 158.22 | 17.79 | 2.80 | 0.34 | 1.64 | 0.19 | | | | | | |
| 49.9 | 0.9 | 144.94 | 16.31 | | | 2.01 | 0.23 | | | | | | |
| 48.2 | 1.0 | 116.88 | 13.15 | | | 2.34 | 0.26 | | | | | | |
| 46.2 | 1.1 | 97.87 | 11.04 | | | 2.92 | 0.33 | | | | | | |
| 44.1 | 1.2 | 57.41 | 6.52 | | | 2.70 | 0.31 | | | | | | |
| 41.9 | 1.3 | 20.16 | 2.36 | | | 0.71 | 0.08 | | | | | | |
| 39.6 | 1.5 | 2.92 | 0.72 | | | 2.53 | 0.29 | | | | | | |
| 37.2 | 1.6 | 0.52 | 1.18 | | | 3.12 | 0.35 | | | | | | |
| 34.7 | 1.7 | 1.32 | 0.28 | | | 0.69 | 0.08 | | | | | | |
| 32.0 | 1.8 | 2.58 | 0.45 | | | 1.99 | 0.22 | | | | | | |
| | | | | | | | | | | | | | |